\documentclass[aps,prl,twocolumn,amsmath,showpacs,amssymb,superscriptaddress,nofootinbib]{revtex4}
\usepackage{graphicx}
\usepackage{dcolumn}
\usepackage{bm}
\usepackage{amssymb,amsmath}
\usepackage{latexsym}
\usepackage{color}
\usepackage[normalem]{ulem} 
\usepackage{cancel}
\allowdisplaybreaks
\newcommand{\Caltech}{\affiliation{Theoretical Astrophysics 350-17,
    California Institute of Technology, Pasadena, CA 91125}}
\newcommand{\Cornell}{\affiliation{Center for Radiophysics and Space
    Research, Cornell University, Ithaca, New York, 14853}}
\newcommand{\NITHEP}{\affiliation{National Institute of Theoretical Physics, Private Bag X1 Matieland, Stellenbosch, South Africa, 7602}}
\newcommand{\Stias} {\affiliation{
Stellenbosch Institute for Advanced Study (STIAS),
Wallenberg Research Centre at Stellenbosch University,
Marais Street,
Stellenbosch 7600,
South Africa}}
\usepackage{notoccite}
\newcommand{\be}{\begin{equation}}
\newcommand{\ee}{\end{equation}}
\newcommand{\ba}{\begin{eqnarray}}
\newcommand{\ea}{\end{eqnarray}}

\begin{document}
\title{Frame-Dragging Vortexes  and Tidal Tendexes
Attached to Colliding Black Holes: \\
Visualizing the Curvature of Spacetime}
\author{Robert Owen} \Cornell
\author{Jeandrew Brink} \NITHEP
\author{Yanbei Chen} \Caltech
\author{Jeffrey D. Kaplan} \Caltech
\author{Geoffrey Lovelace} \Cornell
\author{Keith D. Matthews} \Caltech
\author{David A. Nichols} \Caltech
\author{Mark Scheel} \Caltech
\author{Fan Zhang} \Caltech
\author{Aaron Zimmerman} \Caltech
\author{Kip S.\ Thorne} \Caltech \Stias

\begin{abstract}
When one splits spacetime into space plus time, 
the spacetime curvature (Weyl tensor) gets 
split into an ``electric'' part 
$\mathcal E_{jk}$ that describes tidal gravity 
and a ``magnetic'' part $\mathcal B_{jk}$ that describes differential 
dragging of inertial frames.  We introduce tools for visualizing 
$\mathcal B_{jk}$ (frame-drag vortex lines, their vorticity,
and vortexes) and  
$\mathcal E_{jk}$ (tidal tendex lines, their tendicity, and tendexes),
and also visualizations of a black-hole horizon's (scalar) vorticity and 
tendicity. 
We use these
tools to elucidate the nonlinear dynamics of curved spacetime in merging
black-hole binaries.  

\end{abstract}

%04.25.dg = Binary black holes
%04.25.D- = Numerical relativity
%04.30.-w = gravitational waves
%04.25.Nx = Post-Newtonian, perturbation theory, related approximations
\pacs{04.25.dg, 04.25.D-, 04.30.-w, 04.25.Nx}

\maketitle

{\it Introduction.}---
When one foliates spacetime with 
spacelike hypersurfaces, 
the Weyl curvature tensor $C_{\alpha\beta\gamma\delta}$
(same as Riemann in vacuum) splits into 
``electric'' and ``magnetic'' parts 
$\mathcal E_{jk} = C_{\hat 0 j \hat 0 k}$ and 
$\mathcal B_{jk} = \frac12 \epsilon_{jpq} {C^{pq}}_{k\hat0}$  
(see e.g.\ \cite{Maartens1998} and references therein);
both $\mathcal E_{jk}$ and $\mathcal B_{jk}$ 
are spatial, symmetric, and trace-free.  
Here the indices are in the reference frame 
of ``orthogonal observers'' who move orthogonal to the space slices; 
$\hat 0$ is their time component, 
$\epsilon_{jpq}$ is their spatial Levi-Civita tensor, and throughout 
we use units with $c=G=1$.  

Because two orthogonal observers separated by a tiny spatial vector $\bm \xi$ experience a
relative tidal acceleration 
$\Delta a_j = - \mathcal E_{jk}\xi^k$, 
$\mathcal E_{jk}$ is called the \textit{tidal field}. 
And because a gyroscope at the tip of $\bm \xi$ precesses due to frame dragging 
with an angular velocity 
$\Delta\Omega_j = \mathcal B_{jk} \xi^k$ relative to inertial frames 
at the tail of $\bm \xi$, 
we  
call $\mathcal B_{jk}$ the {\it frame-drag field.} 

\textit{Vortexes and Tendexes 
in Black-Hole Horizons.}---For a binary black hole, 
our
space slices intersect the 3-dimensional (3D) event horizon in a 2D horizon
with inward unit normal $\mathbf N$;
so $\mathcal B_{NN}$ is the 
rate the frame-drag angular velocity around $\mathbf N$ increases
as one moves inward through the horizon. 
Because of the connection between 
rotation and vorticity, 
we call $\mathcal B_{NN}$ the horizon's
\textit{frame-drag vorticity}, or simply its \textit{vorticity}. 

Because $\mathcal B_{NN}$ is boost-invariant
along $\mathbf N$ \cite{Thorne-Price:1986kipversion}, 
the horizon's vorticity is independent of how fast the orthogonal
observers 
fall through the horizon, and is even unchanged if the observers hover immediately above the horizon 
(the
FIDOs of the ``black-hole membrane paradigm'' 
\cite{Thorne-Price-MacDonald:Kipversion}).

\begin{figure}[b]
\includegraphics[width=0.95\columnwidth]{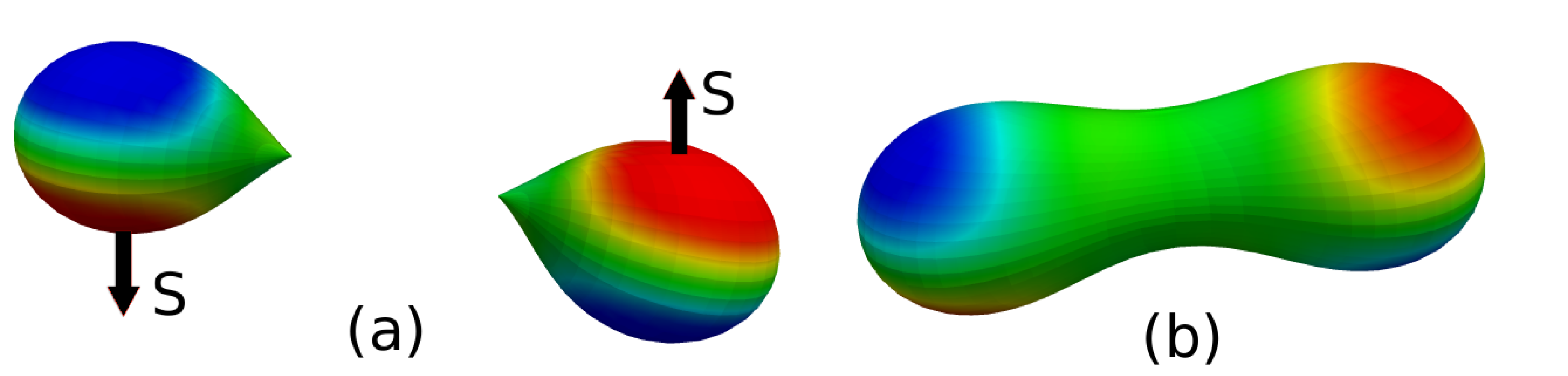}
\caption{ 
Vortexes (with positive vorticity blue, 
negative vorticity red) on the 2D event horizons of 
spinning, colliding black holes, just before and just after 
merger. (From the simulation reported in~\cite{Lovelace:2009}.)}
\label{fig:HO-horizonvortexes}
\end{figure}

Figure \ref{fig:HO-horizonvortexes} shows snapshots of the horizon 
for two identical black holes with transverse, oppositely directed spins 
$\mathbf S$, 
colliding head on. 
Before the collision, each horizon has a negative-vorticity region
(red) centered on $\mathbf S$, and a positive-vorticity region (blue)
on the other side.
We call these regions of concentrated vorticity 
\textit{horizon vortexes}.  
Our numerical simulation \cite{Lovelace:2009} shows the four vortexes being transferred 
to the merged horizon 
(Fig.\ \ref{fig:HO-horizonvortexes}b), then retaining their identities, but
sloshing between positive and negative vorticity and gradually dying, as the hole settles into its final Schwarzschild state; see the movie in 
Ref.~\cite{Headon05aaMovie}.

Because
$\mathcal E_{NN}$ measures the strength of the tidal-stretching
acceleration felt by orthogonal observers as they fall through (or hover above)
the horizon, we call it the horizon's {\it tendicity}  
(a word coined by David Nichols from the Latin {\it tendere}, ``to stretch'').
%cf.\ {\it vorticity} from {\it vertere}, ``to turn''.) 
On the two
ends of the merged horizon in Fig.\ \ref{fig:HO-horizonvortexes}b
there are regions of strongly enhanced tendicity, called {\it tendexes};
cf.\ Fig.\ \ref{fig:SuperkickHorizonPhase} below.

An orthogonal observer falling through the 
horizon carries an orthonormal
tetrad consisting of her 4-velocity $\bf U$, the
horizon's inward normal $\bf N$,
and transverse vectors 
$\bf e_2$ and $\bf e_3$. 
In the null tetrad 
${\bf l} = ({\bf U} - {\bf N})/\sqrt 2$ (tangent to horizon generators),
${\bf n} = ({\bf U} + {\bf N})/\sqrt 2$, 
${\bf  m} = ({\bf e_2} + i {\bf e_3})/\sqrt 2$, and
${\bf m}^*$, 
the Newman-Penrose Weyl scalar $\Psi_2$~\cite{Newman1962} is
$\Psi_2 =( \mathcal E_{NN} + i\mathcal B_{NN})/2$. 
Here we use sign conventions of \cite{FrolovNovikov1998kip}, appropriate 
for our (-\,+++) signature. 

\begin{figure}[t!]
\includegraphics[width=0.94\columnwidth]{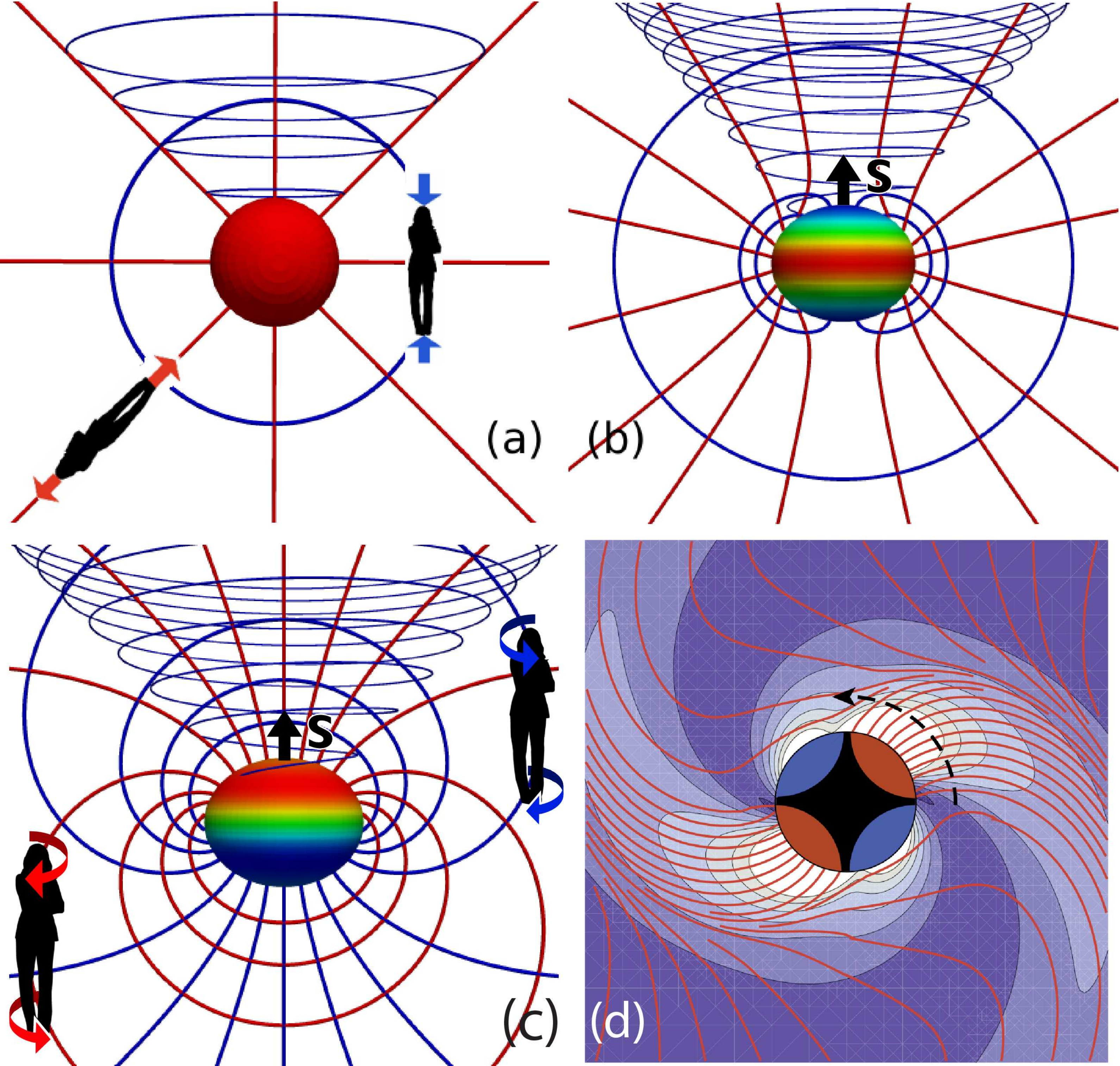}
\caption{
Four different black holes, with horizons colored by their tendicity (upper 
two panels) or vorticity (lower two panels), ranging from most negative (red)
to most positive (blue); and with a Kerr-Schild horizon-penetrating foliation
%
%
%  Problem: This figure contains a citation that does not appear in the
%  main text.  Latex will number this citation in the order in which it
%  appears in the document.  However, PRL style is that this citation should
%  be numbered *after* the citations that appear in the text, but *before*
%  the citations that appear in the acknowledgements. I cannot figure out
%  how to get latex to do this, so I'm resorting to the following crude hack:
%  I'm *by hand* replacing the citation below with the number 18, and then
%  at the beginning of the acknowledgments I put a \nocite{MTW} (which makes
%  MTW appear as reference 18 in the bibliography).
%
%  -- Mark Scheel, Mar 31 2011
%
%%(Exercise 33.8 of \protect\cite{MTW}).
(Exercise 33.8 of Ref.~[18]).
(a) A nonrotating black hole and its tendex lines; 
negative-tendicity lines are red, and positive blue.  
%tangent planes to constant radius spheres are degenerate eigenvector subspaces.
(b) A rapidly rotating (Kerr) black hole, with spin 
$a/M=0.95$, and its tendex lines. 
(c) The same Kerr black hole and its vortex lines. 
%(Kerr hole linearized in $a/M$).  
(d)  Equatorial plane of a nonrotating black hole that 
is oscillating in an 
odd-parity $l=m=2$ 
quasinormal mode, with negative-vorticity 
vortex lines emerging from red horizon vortexes. The lines' vorticities are 
indicated by contours and colors; 
the contour lines, in units $(2M)^{-2}$ and going outward from the hole,
are -10, -8, -6, -4, -2.
}
\label{fig:SchPerts}
\end{figure}

Penrose and Rindler \cite{Penrose1992} define a complex scalar curvature
$\mathcal K = \mathcal R/4 + i\mathcal X/4$ of the 2D horizon, with 
$\mathcal R$ its intrinsic (Ricci) scalar curvature 
(which characterizes the horizon's shape) and
$\mathcal X$ 
proportional to the 
2D curl of its H\'aj\'i\v{c}ek field \cite{Damour1982} (the space-time part of
the 3D horizon's extrinsic curvature).  Penrose and Rindler show
that $\mathcal K = -\Psi_2 +\mu\rho - \lambda\sigma$, where 
$\rho$, $\sigma$, $\mu$, and $\lambda$ are 
spin coefficients
related to the expansion and shear of the
null vectors $\mathbf l$ and $\mathbf n$, respectively.
In the limit of a shear- and expansion-free horizon (e.g.\ a quiescent
black hole; 
%described by the Kerr metric; 
Fig.\ \ref{fig:SchPerts}a,b,c), 
$\mu\rho - \lambda\sigma$ vanishes, 
so $\mathcal K = -\Psi_2$, whence
$\mathcal R = - 2\mathcal E_{NN}$ and
$\mathcal X = - 2\mathcal B_{NN}$. 
As the dimensionless spin parameter $a/M$
of a quiescent (Kerr) black hole is increased, the scalar curvature $\mathcal R
= - 2 \mathcal E_{NN}$
at its poles decreases, becoming negative for $a/M > \sqrt{3}/2$; see
the blue spots on the poles in Fig.\ \ref{fig:SchPerts}b compared to solid red
for the nonrotating hole in Fig.\ \ref{fig:SchPerts}a.  
In our binary-black-hole simulations, the contributions of the spin 
coefficients 
to $\mathcal K$ on the apparent horizons 
are small [$L2$-norm $\lesssim 1\%$] so
$\mathcal R \simeq - 2\mathcal E_{NN}$ and 
$\mathcal X \simeq - 2\mathcal B_{NN}$, 
except for a time interval
$\sim 5 M_{\rm tot}$ near merger. 
Here $M_{\rm tot}$ is the 
binary's total mass. On the event horizon, the duration of 
spin-coefficient contributions $>1\%$ is somewhat longer, but we do
not yet have a good measure of it.

Because $\mathcal X$ is the 2D curl of a 2D vector, its integral over the 
2D horizon vanishes. Therefore, positive-vorticity regions must be 
balanced by negative-vorticity regions; it is impossible to have a horizon
with just one vortex.  By contrast, 
the Gauss-Bonnet theorem says the
integral of $\mathcal R$ over the 2D horizon is $8\pi$ (assuming 
%the horizon has 
$S_2$ topology), which implies the horizon tendicity $\mathcal E_{NN}$ is 
predominantly negative (because $\mathcal E_{NN} \simeq - \mathcal R / 2$ 
%as argued above, 
and $\mathcal R$ is predominantly positive).  Many black holes 
have negative horizon tendicity
everywhere (an exception is Fig.\ \ref{fig:SchPerts}b), so
their horizon tendexes must be distinguished by deviations of
$\mathcal E_{NN}$ from a horizon-averaged value. 

{\it 3D vortex and tendex lines}---The frame-drag field $\mathcal B_{jk}$ is
symmetric and trace free and therefore is fully characterized by its three 
orthonormal 
eigenvectors 
$\mathbf e_{\tilde j}$ 
%(with $j=1,2,3$)
and their eigenvalues $\mathcal B_{\tilde 1\tilde 1}$, 
$\mathcal B_{\tilde 2\tilde2}$ and $\mathcal B_{\tilde3\tilde3}$.  We call 
the integral curves along $\mathbf e_{\tilde j}$ \textit{vortex lines}, 
and their eigenvalue 
$\mathcal B_{\tilde j\tilde j}$ those lines' 
\textit{vorticity}, and we call a concentration of vortex lines
with large vorticity a {\it vortex}. For the tidal field $\mathcal E_{jk}$
the analogous quantities are {\it tendex lines}, {\it tendicity} 
and {\it tendexes}.  For a nonrotating
(Schwarzschild) black hole, we show a few tendex lines 
in Fig.\ \ref{fig:SchPerts}a; and
for a rapidly-spinning black hole (Kerr metric with $a/M=0.95$) we show
tendex lines in Fig.\ \ref{fig:SchPerts}b and vortex lines in Fig.\ 
\ref{fig:SchPerts}c. 

If a person's body (with length $\ell$) is oriented 
along a positive-tendicity
tendex line (blue in Fig.\ \ref{fig:SchPerts}a), she feels a head-to-foot 
compressional acceleration 
$\Delta a = |\hbox{tendicity}|\ell$; for negative tendicity (red)
it is a stretch.  If her body is oriented 
along a positive-vorticity vortex line (blue in Fig.\ \ref{fig:SchPerts}c), 
her head sees a gyroscope
at her feet precess 
%counterclockwise 
clockwise
with angular speed
$\Delta \Omega = |\hbox{vorticity}|\ell$, and her feet see a gyroscope
at her head also precess 
%counter
clockwise 
at the same rate. 
For negative vorticity 
(red) the precessions are 
%clockwise.
counterclockwise.

For a nonrotating black hole, the stretching tendex lines are radial,
and the squeezing %tendex lines 
ones lie on spheres 
%around the hole 
(Fig.\ \ref{fig:SchPerts}a).  When the hole is spun up to $a/M=0.95$
(Fig.\ \ref{fig:SchPerts}b), its toroidal tendex lines acquire a spiral, 
and its poloidal tendex lines, when emerging from one polar region, 
return to the other polar region. For any
spinning Kerr hole (e.g.\ Fig.\ \ref{fig:SchPerts}c), the vortex 
lines from each polar region reach around the hole and return to the
same region. The red vortex lines
from the red north polar region constitute a 
%{\it clockwise vortex};
{\it counterclockwise vortex}:
the blue ones from the south polar region constitute a 
%{\it counterclockwise vortex}. 
{\it clockwise vortex}.

As a dynamical example, consider a Schwarzschild black hole's 
fundamental odd-parity $l=m=2$ quasinormal mode of pulsation,
which is governed by Regge-Wheeler perturbation 
theory~\cite{ChandraDetweiler1975:kip} and has
angular eigenfrequency 
$\omega = (0.74734-0.17792i)/2M$, with $M$ the hole's mass.  From
the perturbation equations, we have deduced the 
mode's
horizon vorticity:  
$\mathcal B_{NN} = \Re \{ 9 \sin^2\theta
/(2i\omega M^3) \exp[2i\phi-i\omega(\tilde t+2M)]\}$. 
(Here $\tilde t$ is the ingoing Eddington-Finklestein time coordinate,
and the mode's Regge-Wheeler radial eigenfunction
$Q(r)$ is normalized to unity near the horizon.)
At time $\tilde t=0$, this $\mathcal B_{NN}$ exhibits
four horizon vortexes 
[red and blue in Fig.\ \ref{fig:SchPerts}d],
centered on the equator at $(\theta,\phi) = (\pi/2, 1.159+k\pi/2)$ ($k=0,1,2,3$), and 
with central vorticities 
$\mathcal B_{NN} = - (-1)^k  39.22 /(2M)^2$.  
From  
analytic formulae for $\mathcal B_{jk}$ and a numerical $Q(r)$, 
we have deduced the equatorial-plane red vortex lines and vorticities 
shown in Fig.\ \ref{fig:SchPerts}d. As time $\tilde t$ passes, the vortexes rotate
counterclockwise, so they resemble water splayed out from a turning sprinkler.
The transition from near zone to wave zone is at $r \sim 4M$ (near the outermost part
of the second contour line). 
As one moves into the wave zone, each of the red vortexes is 
smoothly transformed into a gravitational-wave trough 
%(transverse vortex lines with
%negative vorticity),
and the 3D vortexes that
emerge from the blue horizon vortexes (concentrated in the dark
region of this figure) are transformed into gravitational-wave crests. 

%{\it Relation to frame dragging in linearized theory.}---In the linearized approximation
%to general relativity, it is conventional to describe a spinning body's
%frame dragging by gyroscopes' angular velocity of precession relative 
%to inertial frames at ``infinity'': 
%$
%\bm \Omega=  
%  \left[\bm 3 (\bm S\cdot \hat{\bm r})\hat{\bm r} -\bm S\right]/r^3 \, 
%\label{OmegaLT}
%$
%(Eq.\ (40.37) of \cite{MTW}).
%This is minus half the body's
%``gravitomagnetic 
%field'' and it has dipolar stream lines. 
%Here $\bm S$ is the body's angular momentum, 
%$r$ is distance from its center and $\hat{\bm r}$ is the unit radial vector.   
%When gravity is strong and dynamical (as for colliding black holes), there is no 
%unique way to compare gyroscopic pointing with inertial frames 
%at infinity, so   
%we must focus on differential frame dragging as embodied in $\mathcal B_{jk}$.  In linearized theory, $\mathcal B_{jk}$ is the spatial gradient of $\bm \Omega$ 
%(Eq.\ (5.45b) of \cite{Thorne-Price-MacDonald:Kipversion}),
%which for our spinning body is
%$
%\mathcal B_{jk} = \left[ 6 S_{(j}{\hat r}_{k)} + 3 (\bm S\cdot \hat{\bm r}) 
%\delta_{jk} - 15 (\bm S \cdot
%\hat{\bm r}) {\hat r}_j {\hat r}_k\right]/r^4$ (where subscript parentheses 
%mean symmetrize).  
%This $\mathcal B_{jk}$ 
%has vortex lines 
%like those of a spinning black hole 
%(Fig.\ \ref{fig:SchPerts}a). 

{\it Vortex and Tendex Evolutions in Binary Black Holes (BBHs).}--- We 
have explored the evolution of frame-drag
vortexes and tidal tendexes in numerical simulations of three BBHs 
%configurations 
that differ greatly from each other.  
%We shall discuss them in turn.

\begin{figure}[t!]
\includegraphics[width=0.95\columnwidth]{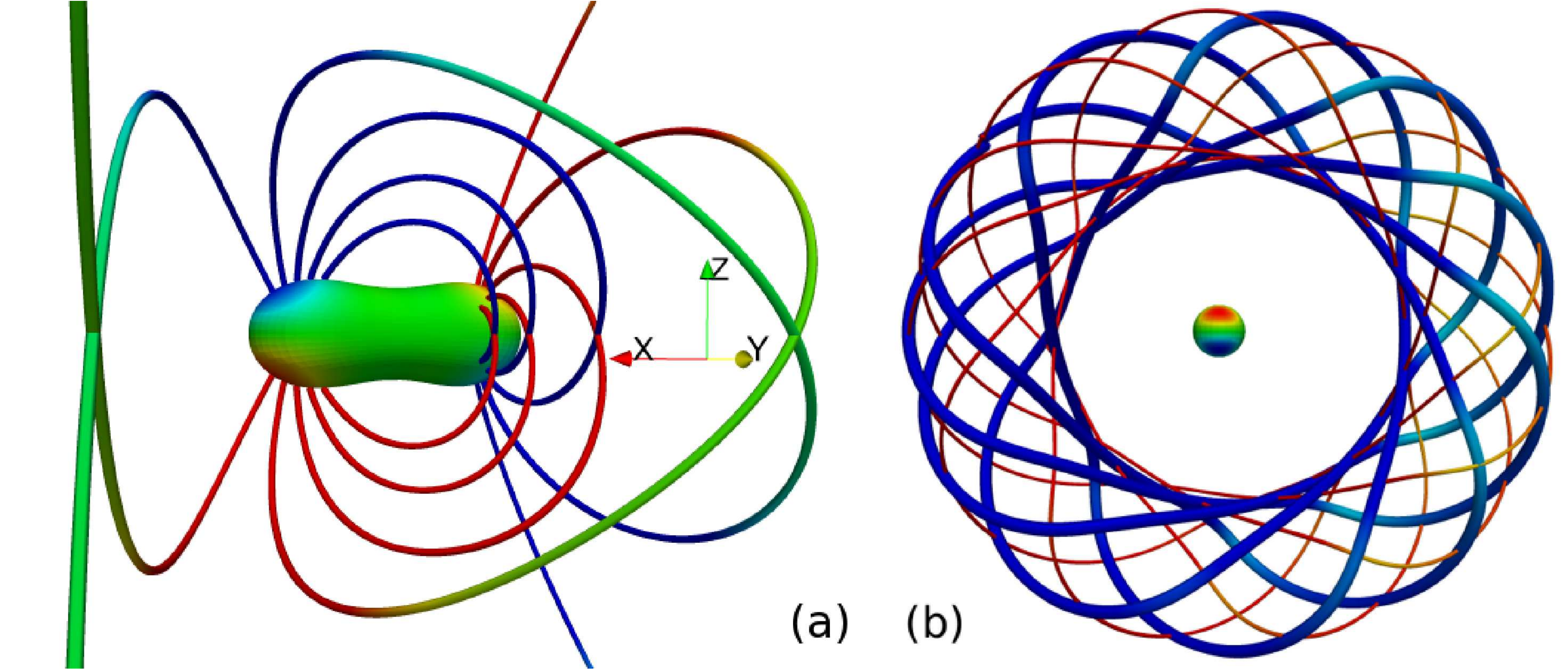}
\caption{Head-on, transverse-spin simulation: (a) 
Shortly after merger, vortex lines link horizon vortexes of same
polarity (red to red; blue to blue). Lines are color coded by vorticity
(different scale from horizon).
(b) Sloshing of near-zone vortexes
generates vortex loops traveling outward as gravitational waves; 
thick and thin lines are orthogonal vortex lines. 
}
%thick and thin lines are the two transverse vortex
%lines color coded by their vorticity (different scale than horizon vorticity). 
%}
\label{fig:HO-vortexLines}
\end{figure}

Our first simulation (documented in Ref.~\cite{Lovelace:2009}; movies in
Ref.~\cite{Headon05aaMovie}) is 
the head-on, transverse-spin merger depicted in 
Fig.\ \ref{fig:HO-horizonvortexes} above, with spin magnitudes $a/M=0.5$.
As the holes approach each other then merge, their 3D vortex lines, 
which originally link a horizon vortex to itself on a single hole 
(Fig.\ \ref{fig:SchPerts}c),
reconnect so on the merged hole they link one horizon vortex to the other
of the same polarity (Fig.\ \ref{fig:HO-vortexLines}a).  After merger, the
near-zone 3D vortexes slosh (their vorticity oscillates between positive and
negative), generating vortex loops (Fig.\ \ref{fig:HO-vortexLines}b)
that travel outward as gravitational waves.  
%The blue and red 
%transverse (nonradial) 
%loop segments
%constitute a gravitational-wave crest. 

\begin{figure} [t!]
\includegraphics[width=3.2in]{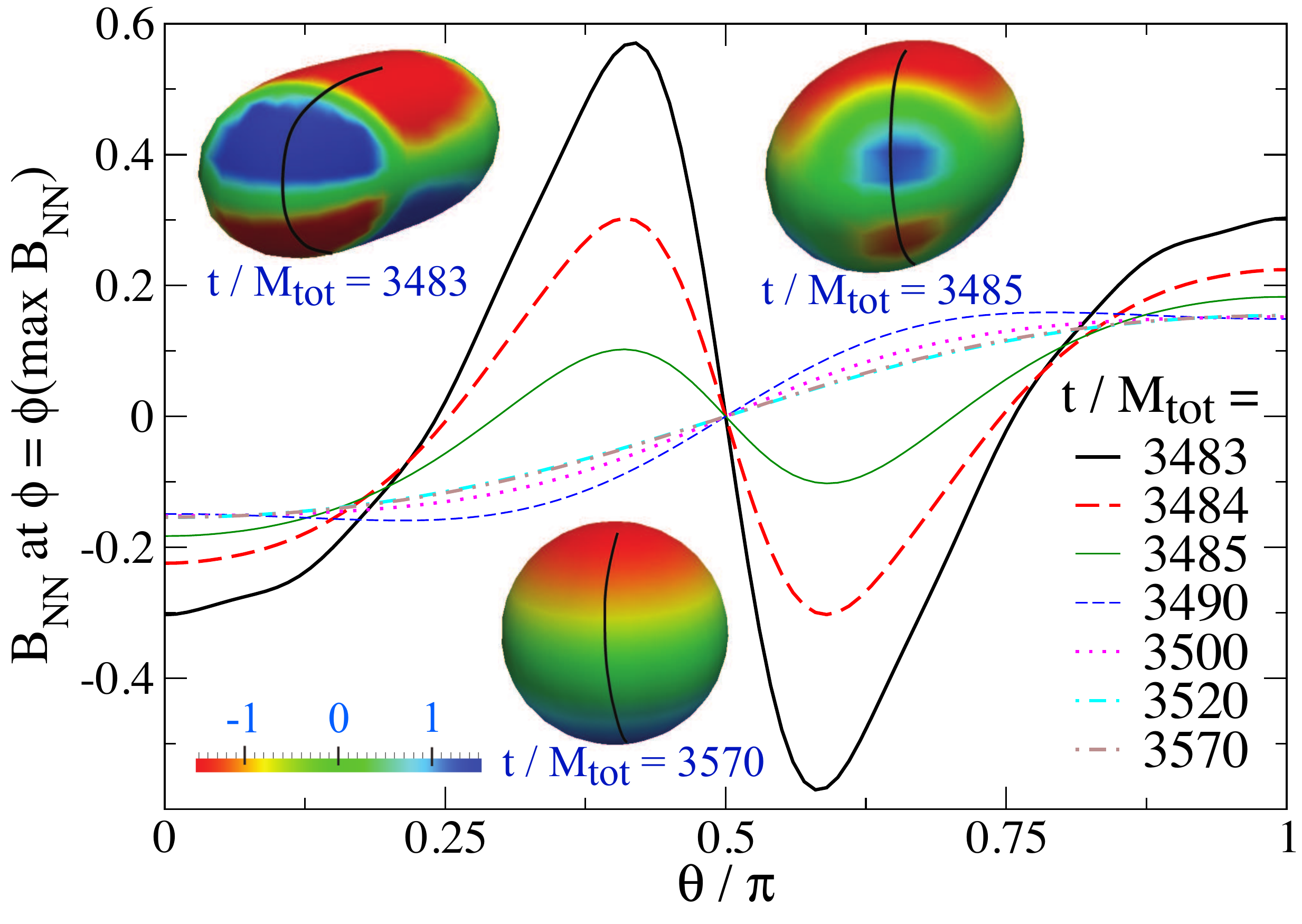}
\caption{
Insets: snapshots of the common apparent horizon for the
$a/M=0.95$ anti-aligned simulation, color coded with 
the horizon vorticity
$B_{NN}$. 
%(in arbitrary units).  
Graphs: 
%the horizon vorticity 
$B_{NN}$ as a function of polar angle 
$\theta$ at the azimuthal angle $\phi$ 
that bisects the four vortexes (along the black curves in snapshots). 
\label{fig:Anti95}}
\end{figure}

Our second simulation (documented in Ref.~\cite{Lovelace2010};
movies in Ref.~\cite{Inspiral95aaMovie})
is the inspiral and merger of two identical, fast-spinning 
holes ($a/M=0.95$) with spins 
antialigned to the 
orbital angular momentum. 
Figure \ref{fig:Anti95} 
shows the evolution of the vorticity $\mathcal B_{NN}$
on the common apparent horizon beginning just after merger 
(at time $t/M_{\rm tot}=3483$),
as seen in a frame that co-rotates with the small horizon vortexes.  
In that frame, the small vortexes 
(which arise from the initial holes' spins)
appear to diffuse into the two large central vortexes (which arise from
the initial holes' orbital angular momentum), annihilating some of their
vorticity. (This is similar to the diffusion and annihilation of
magnetic field lines with opposite polarity threading 
a horizon~\cite{Thorne-Price-MacDonald:Kipversion}.)   
Making this heuristic
description quantitative, or disproving it, is an important challenge.    
  
Our third simulation (see movies in 
Ref.~\cite{ExtremeKickVortexMovie})
is a variant of the ``extreme-kick'' merger
studied by Campanelli et\ al.\ \cite{Campanelli2007} 
and others \cite{Gonzalez2007b,LoustoZlochower:2010}:
two identical holes, merging from an initially circular orbit,
with oppositely directed spins $a/M=0.5$ 
lying in the orbital $(x,y)$ plane. 
In this case, the vortexes and tendexes in the merged 
hole's $(x,y)$ plane rotate 
as shown in Fig.~\ref{fig:SchPerts}d.
We have tuned the initial conditions to make the final hole's kick (nearly)
maximal, in the $+z$ direction.  
The following considerations explain the %physical
origin of this maximized kick:   

In a plane gravitational wave, all the vortex and tendex lines with nonzero eigenvalues
lie in the wave fronts and make angles of 45 degrees to each
other (bottom inset of Fig.\ \ref{fig:SuperkickHorizonPhase}.) For vectors $\bm E$ (parallel to solid, 
positive-tendicity tendex line)  
and $\bm B$ (parallel to dashed, positive-vorticity vortex line), $\bm E \times \bm B$
is in the wave's propagation direction.   

Now, during and after merger, the black hole's near-zone rotating tendex lines 
(top left inset in
Fig.\ \ref{fig:SuperkickHorizonPhase}) acquire accompanying vortex lines as they 
travel outward 
into the wave zone 
%(Fig.\ \ref{fig:SchPerts}b) 
and become gravitational waves; 
and 
%similarly 
the rotating near-zone vortex lines acquire accompanying tendex lines.  
Because of the evolution-equation duality between 
$\mathcal E_{ij}$ and $\mathcal B_{ij}$,
the details of this wave formation are essentially the same for the rotating
tendex and vortex lines.  Now, in the near zone,
the vectors $\bm E$ and $\bm B$ along the tendex and vortex 
lines (Fig.\ \ref{fig:SuperkickHorizonPhase})
make the same angle with respect to each other as in a gravitational wave (45 degrees) and
have $\bm E \times \bm B$ in the $-z$ direction. 
This means that 
%the superposition of weak gravitational waves consistent with 
%this configuration of vortex and tendex lines is one that carries momentum 
%predominantly in the $-z$ direction, with a counterbalancing recoil of the 
%black hole in the $+z$ direction.  
%If the angle between near-zone vortex and tendex 
%lines were different, then some component of gravitational waves propagating 
%in the $+z$ direction would be needed to attain that angle, meaning that some 
%of the field momentum is balanced by field momentum in the other direction, 
%leaving less momentum remaining to balance a black hole kick.
%This means that the gravitational waves 
%they generate will superpose constructively in the $-z$ direction and destructively
%in the $+z$ direction, maximizing the gravitational-wave momentum flow in the
%$-z$ direction and maximizing the black-hole kick in $+z$. 
%%% KIP'S NEW PHRASING:
the gravitational waves produced by the rotating near-zone tendex lines and those
produced by the rotating near-zone vortex lines will superpose 
constructively
in the $-z$ direction and destructively in the $+z$ direction, leading
to a maximized gravitational-wave momentum flow in the $-z$ direction
and maximized black-hole kick in the $+z$ direction.
%%%
An extension of this reasoning shows that the black-hole kick velocity
is sinusoidal in twice the angle between the merged hole's near-zone rotating vortexes
and tendexes, in accord with simulations. 
 
\begin{figure} [t!]
\includegraphics[width=0.95\columnwidth]{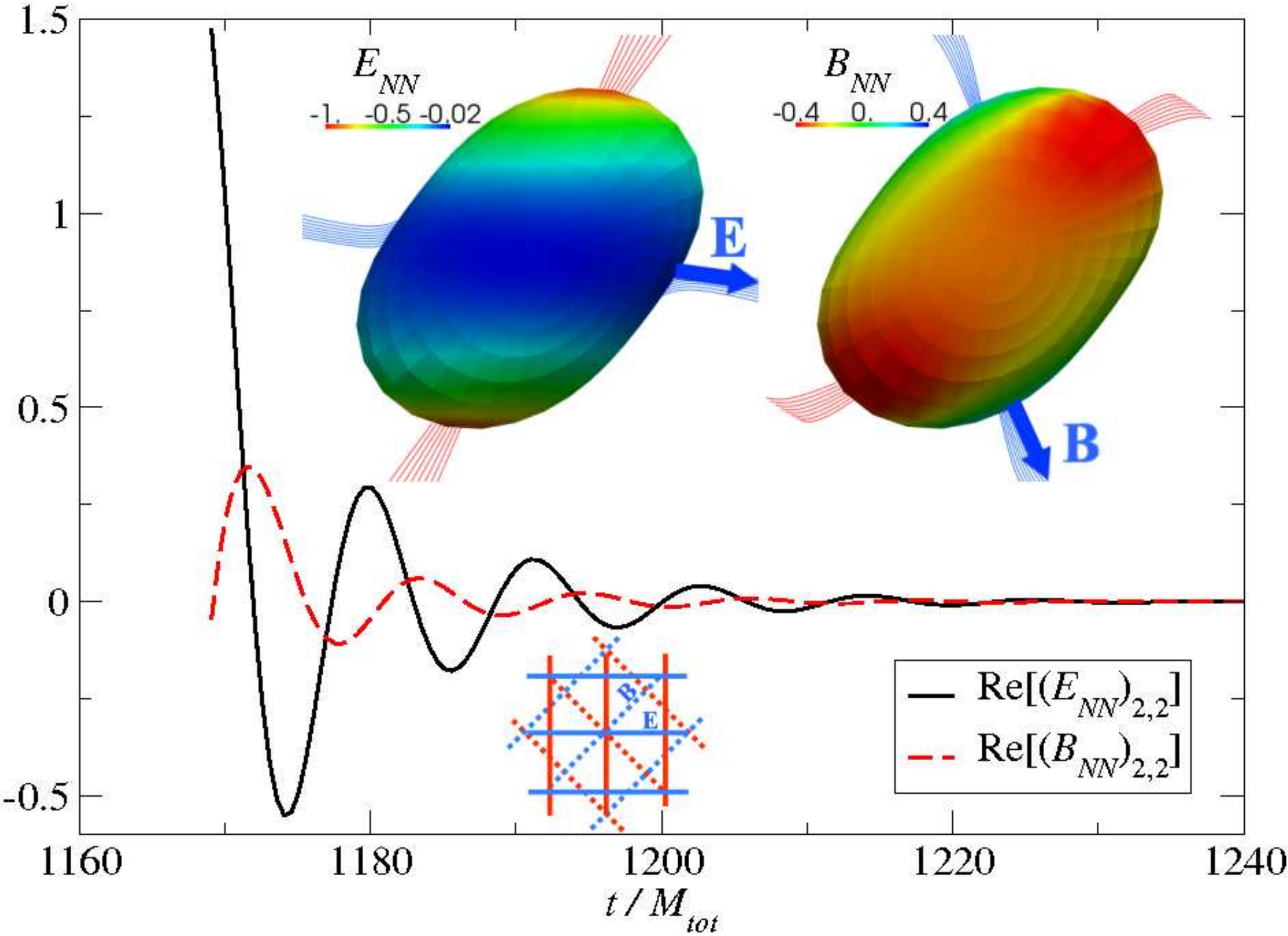}
\caption{
Bottom inset: tendex and vortex lines
for a plane gravitational wave; 
%propagating out of the paper or screen; 
$\bm E \times \bm B$ is in the propagation direction.  
Upper two  insets: for the ``extreme-kick simulation'',
as seen looking down the merged hole's rotation axis ($-z$ direction): 
the 
%merged hole's 
apparent horizon 
color coded with the horizon tendicity (left inset) and vorticity
(right inset), and with 3D vortex lines and tendex lines emerging from
the horizon.  The tendexes with the most positive tendicity (blue; $\bm E$) lead the 
positive-vorticity vortexes (blue, $\bm B$) by about $45^{\rm o}$ as they rotate
counterclockwise.  This $45^{\rm o}$ lead is verified in the oscillating
curves, which show the rotating 
$\mathcal B_{NN}$ and $\mathcal E_{NN}$
projected onto a nonrotating 
$\ell=2$, $m=2$ spherical harmonic.  
}
\label{fig:SuperkickHorizonPhase}
\end{figure}

{\it Conclusions.}---In our BBH simulations, the nonlinear dynamics of curved spacetime appears to be dominated by (i) the transfer
of spin-induced frame-drag vortexes from the initial holes to the final merged hole, (ii) the creation
of two large vortexes on the merged hole associated with the orbital angular momentum,
(iii) the subsequent sloshing, diffusion, and/or rotational motion 
of the spin-induced vortexes,
(iv) the formation of strong negative $\mathcal{E}_{NN}$ poloidal tendexes on the merged horizon at the locations of the original two holes, associated with the horizon's elongation, and a positive $\mathcal{E}_{NN}$ tendex at the neck where merger occurs, 
%whose tendex lines radiate outward perpendicular to the horizon in a disk, 
and (v) the oscillation, diffusion, and/or circulatory
motion of these tendexes.  

We {\it conjecture} that there is no other important dynamics in the 
merger and ringdown
of BBHs.  If so, there are important 
consequences:
(i) This could account for the surprising simplicity of the BBH 
gravitational waveforms predicted by simulations. 
(ii) A systematic study of frame-drag vortexes and tidal tendexes in
BBH simulations may produce improved understanding of BBHs, including their
waveforms and kicks. 
The new waveform insights may 
lead to improved functional forms for waveforms that are tuned via
simulations 
to serve as templates in LIGO/VIRGO data analysis. 
(iii) Approximation techniques that aim to smoothly cover the
full spacetime of BBH mergers (e.g.\ the combined Post-Newtonian and
black-hole-perturbation theory method \cite{Nichols2010}) might 
be made to capture
accurately the structure and dynamics of frame-drag vortexes and tidal 
tendexes.  If so, these approximations may become powerful 
and accurate tools for generating BBH
waveforms.

%% The precise forms of the vortex lines, tendex lines, and horizon vorticity 
%% and tendicity are gauge dependent, but this dependence is `quarantined' to 
%% the choice of space slices. 
%% %after which they are unique. 
%% As for electric and magnetic fields,
%% this slicing dependence is easily understood in terms of local Lorentz
%% transformations. We expect that for some class of `reasonable' 
%% slicings (including those of robust numerical BBH simulations), the vortex and 
%% tendex behaviors around BBHs will be qualitatively slicing independent.

%
%
% The \nocite{MTW} below is my ugly hack to a reference numbering problem.
% Please see the larger comment above in the caption to Figure 2 for
% an explanation.
%
% -- Mark Scheel
%
%
{\it Acknowledgements.}---\nocite{MTW}We thank Larry Kidder and
Saul Teukolsky for helpful discussions.  Our simulations have been
performed using the Spectral Einstein Code ({\tt SpEC})~\cite{SpECwebsite}.
This research was supported by
NSF grants PHY-0601459, PHY-0653653, PHY-0960291,
PHY-0969111, PHY-1005426, and CAREER grant PHY-0956189,
by NASA grants NNX09AF97G and NNX09AF96G,
and by  the Sherman Fairchild Foundation, the Brinson Foundation,
and the David and Barbara Groce fund.

\bibliography{References/References}

\end{document}